\documentclass[namedreferences,hyperref,optionalrh]{spr-sola}
\usepackage{graphicx}        
\usepackage{color,amsmath,amssymb}           
\usepackage{natbib}
\setcitestyle{authoryear, maxnames=3}

\usepackage{multirow}

\chardef\us=`\_
\newcommand{\kms}{\,km\,s$^{-1}$}
\newcommand{\Fm}{${\rm F_{CME,m}}$}
\newcommand{\Ncme}{${\rm N_{CME}}$}
\newcommand{\vcme}{${\rm V_{CME}}$}
\newcommand{\acme}{${\rm a_{CME}}$}
\newcommand{\Fcme}{${\rm F_{CME}}$}

\newcommand{\vcmem}{${\rm V_{CME,m}}$}
\newcommand{\ardt}{${\rm \Delta T_{A}}$}
\newcommand{\vsp}{${\rm V^{sp}_{CME}}$}
\newcommand{\rcme}{${\rm R_{CME}}$}
\newcommand{\rcmeb}{${\rm R_{CME,b}}$}
\newcommand{\rcmef}{${\rm R_{CME,f}}$}
\newcommand{\twf}{$\mathrm{t_{w,f}}$}
\newcommand{\twb}{$\mathrm{t_{w,b}}$}
\newcommand{\vcmerb}{${\rm V_{CME,r,b}}$}
\newcommand{\vcmerf}{${\rm V_{CME,r,f}}$}
\newcommand{\lcf}{${\rm l_{int,f}}$}
\newcommand{\lcb}{${\rm l_{int,b}}$}
\newcommand{\Rsun}{${\rm R_{\odot}}$}
\newcommand{\msq}{\,m\,s$^{-2}$}
\newcommand{\rs}{$r_{s}$}
\begin{document}

\begin{frontmatter}
\title{Role of CME clusters and CME-CME interactions in producing sustained $\gamma$-ray emission.}

\author[addressref={aff1,aff2},corref,email={atul.mohan@nasa.gov}]{\inits{A.}\fnm{Atul}~\snm{Mohan}\orcid{0000-0002-1571-7931}}
\author[addressref={aff1,aff2},email={pertti.a.makela@nasa.gov}]{\inits{P.}\fnm{Pertti}~\snm{M\"{a}kel\"{a}}\orcid{0000-0002-8182-4559}}
\author[addressref=aff1]{\inits{N.}\fnm{Natchimuthuk}~\snm{Gopalswamy}\orcid{0000-0001-5894-9954}}
\author[addressref={aff1,aff2}]{\inits{S.}\fnm{Sachicko}~\snm{Akiyama}\orcid{0000-0002-7281-1166}}
\author[addressref={aff2,aff1}]{\inits{S.}\fnm{Seiji}~\snm{Yashiro}\orcid{0000-0002-6965-3785}}
\address[id=aff1]{NASA Goddard Space Flight Center,  8800 Greenbelt Road, Greenbelt, MD, 20771, USA}
\address[id=aff2]{The Catholic University of America, Washington, DC, USA}

\runningauthor{Mohan, A. et al.}
\runningtitle{\textit{Role of CME-CME interactions in SGRE production}}

\begin{abstract}
{Fast (\vcme$>$1000\kms) coronal mass ejections (CMEs) capable of accelerating protons beyond 300\,MeV are thought to trigger hours-long sustained $\gamma$-ray emission (SGRE) after the impulsive flare phase.
{ Meanwhile, CME-CME interactions can cause enhanced proton acceleration, increasing the fluxes of solar energetic particles.
This study explores the role of fast CME interactions in SGRE production during CME clusters, which we define as a series of CMEs linked to $>$C-class flares with waiting times $<$1\,day from the same active region (AR).
We focus on clusters in major CME-productive ARs (major ARs), by defining a major AR as one that produced $>$1 CME-associated major ($>$M-class) flare. 
The study identified 76 major ARs between 2011 and 2019, of which 12 produced all SGRE events.}
SGRE-producing ARs exhibit higher median values for the speed of their fastest CMEs (2013 vs. 775\kms) and the class of their strongest flares (X1.8 vs. M5.8), compared to SGRE-lacking ARs. They also produced relatively faster CMEs (median speed: 1418 vs. 1206.5\kms), with the SGRE-associated CMEs occurring during periods of higher CME rates than typical fast CME epochs.
Twelve of 22 (54.5\%) SGRE events and 5 of 7 (71.4\%) long-duration ($>$10\,h) SGRE events occurred during CME clusters, with high chances of CME-CME interactions.
A case study on very active major ARs showed that all SGRE-associated CMEs with \vcme$\lesssim$2000\kms\ underwent CME-CME interactions within 10\,\Rsun, while SGRE-associated CMEs faster than 3000\kms\ did not undergo interactions.
}
\end{abstract}
\end{frontmatter}

\section{Introduction}
     \label{S-Introduction} 
{A sustained $\gamma$-ray emission (SGRE) event is characterized by an extended period of enhanced $\gamma$-ray flux lasting up to several hours, beginning after the impulsive phase of the associated flare.}
SGRE events were first reported and demonstrated to be distinct from the impulsive flare-phase emission by \cite{Forrest85_LatePhGRE_disc} using data from the Gamma Ray Spectrometer \citep[GRS;][]{Forrest80_GRSinstrument} onboard the Solar Maximum Mission.
{Subsequent missions such as Gamma-1~\citep{Akimov91_gamma1tel} and Energetic Gamma-Ray Experiment Telescope~\citep{Kanbach89_EGRET} enabled SGRE spectral observations up to a few GeV.}
{Detailed modeling of various SGRE events highlighted the need for coronal shocks in accelerating protons above 300\,MeV to produce the observed $>$100\,MeV spectrum~\citep{murphy87,Akimov92_SGREdisc_modeling}.
The reported similarity between the spectra of an SGRE and the co-temporal solar energetic particle (SEP) events, likely accelerated by the same shock, is noteworthy~\citep{Forrest85_LatePhGRE_disc}. 
However, the SGRE event sample size remained scarce until the advent of Fermi-Large Area Telescope~\citep[Fermi/LAT;][]{Atwood09_FERMI_LAT} which offered a superior sensitivity and a wider field of view, allowing extended periods of solar monitoring. 
The latest Fermi/LAT SGRE event catalog~\citep{Ajello21_GREcat} has 22 events with durations over 3 hours ($>$3\,h SGRE events).
Various studies have demonstrated the association of $>$3\,h SGRE events with coronal mass ejection (CME) shocks using $\gamma$-ray to radio band data~\citep[e.g.][]{Ackermann17_BTLevents,gopal18_sep17GLE-CME_link,Melissa22_BTL-CMElink,makela23_iniCMEaccel-SGRElink}. 
The SGRE-associated CMEs resemble CMEs that accelerate protons beyond 100\,MeV causing ground-level enhancements, recorded by neutron monitors~\citep{gopal18_sep17GLE-CME_link}

Meanwhile, alternative SGRE models have been proposed. A major alternative model invokes large flare loops that trap and re-accelerate particles for hours post-flare~\citep{Ryan91_loopTurbaccel_SGRE, natalie92_Turbflareloop-SGREmodel,Hudson18_lasso,2019ApJ...879...90D}. However, sustaining post-flare proton acceleration for $>$3\,h imposes stringent conditions on the turbulence within the loops and extended particle acceleration episodes in the active regions (ARs), which are strongly debated~\citep{ramaty98_GRE_overview, kahler18_nosustainedflareaccel_longloop}.  
For instance, \cite{Kambach93_11Jun91_SGRE} attempted to explain the SGRE of 1991 June 11 that lasted more than 8 hours using a turbulent flare-loop model. The authors ended up demonstrating that only turbulence-free loops can efficiently trap the particles for so long, contradicting the fundamental framework of the model, which requires turbulence to re-accelerate particles to energies $>$300\,MeV.
}
Also, SGRE sources observed $\sim$35$^\circ$ behind the limb will require unusually large loops to transport particles and cause precipitation on the front side. However, CME shocks that extend across large scales easily explain particle acceleration and magnetic connectivity to the SGRE source, as demonstrated in various events~\citep{Melissa15_BTL,Ackermann17_BTLevents,gopal20_Sep2014SGRE_CMEsrc,Melissa22_BTL-CMElink}.
{Data-driven modeling of three behind-the-limb SGRE events by \cite{Plotnikov17_CME-SGRElink_BTLev} showed that, in each case, a CME shock producing a co-temporal SEP event could be associated with the SGRE.
\cite{gopal18_SGRE-typeIIlink} reported a correlation between the durations of $>$3\,h SGRE events and the co-temporal decameter hectometric (DH) type II bursts driven by CME shocks, providing another strong support for the CME-shock association of SGRE.
The $>$3\,h SGRE events are generally associated with fast ($>$1000\kms) and wide (mostly halo) CMEs with relatively high initial accelerations~\citep{makela23_iniCMEaccel-SGRElink}.
The work presented here focuses on $>$3\,h SGRE events, due to their strong association with CMEs~\citep{gopal19_SGRE_CMEgrp,makela23_iniCMEaccel-SGRElink}.}

{ 
Meanwhile, several studies have shown that CME-CME interactions during periods of successive CMEs with short waiting times, often referred to as CME clusters~\citep{Ruzmaikin11_CMEclustering,Gomez20_CMEclustering}, are often associated with increased SEP fluxes and geomagnetic disturbances~\citep{gopal04_CME-CMEinter-SEP,pommel18_EUPHORIA_CME-inter,Rodriguez20_CMEcluster-enhancedDST,Koehn22_geoeff_CMEintermodel,Mayank24_cme-cmeinter_SWASTi}.
\cite{gopal01_interactCMEs_DHtypeII} discovered enhanced DH type II radio emission during a CME-CME interaction, signifying an increased particle acceleration efficiency.
Later, various other groups also reported similar cases of enhancement in type II burst emission during CME-CME interaction events~\citep[e.g.,][]{Reiner03_CME-CMEinter,Oliveros12_CME-CMEinter01Aug10,Ding14_CME-CMEinter22May13,Temmer14_CME-CMEinter14Feb11}.
Meanwhile, analyzing the CMEs and ARs associated with SGRE events, \cite{gopal19_SGRE_CMEgrp} found that several events occurred during CME clusters hosted by some major CME-productive ARs (major ARs).
CME-CME interactions in these clusters involving fast CMEs could have played a role in accelerating protons beyond 300\,MeV, through mechanisms similar to those proposed to explain enhanced SEP fluxes~\citep[see,][for a review]{lugaz17_CME-CMEinter_Rev}. These $>$300\,MeV protons could precipitate to deeper layers of the solar atmosphere, producing SGRE through various weak nuclear interactions~\citep{ramaty83_gamraylines_flares,murphy87}.
This work will define major ARs and CME clusters relevant in the context of SGRE events, systematically identify and characterize them using various parameters, and investigate the role of CME clusters and CME-CME interactions in SGRE production.
}

{Section~\ref{data} introduces the data sources, methodology, and definitions used in this work. 
Section~\ref{results} presents the major AR catalog and compares the activity characteristics of SGRE-producing and SGRE-lacking major ARs. The section also discusses the properties of fast CMEs, their epochs, and their association with CME clusters. A case study on the role of CME-CME interactions in triggering SGRE events is also presented.
Section~\ref{conclusion} summarizes our conclusions.}
\vspace{-.5cm}
\section{Data and methodology} \label{data}
The $>$3\,h SGRE events in the Fermi/LAT event lists~\citep{Share18_SGRE-CMElink,Ajello21_GREcat} and the CME database compiled by the Coordinated Data Analysis Workshops (CDAW) Data Center ~\citep[CDAW CME catalog;][]{Yashiro04_LASCOCME_Catalog,gopal09_LASCOCMEcatalog} form our main data source.
{Since we are interested only in strong CME-CME interactions capable of producing $>$300\,MeV protons, the study is focused only on CMEs associated with $>$C-class flares. The flare class is a metric of the peak soft X-ray (SXR) flare flux, which is an indicator of the strength of the reconnection event that propels the CME ~\citep{petsov03_Lx_recflx_solar-stellar,Kazachenko23_ribbonDB,Atul24_modGBR}. The peak SXR flare flux therefore correlates with the CME speed (\vcme) observed in the LASCO C2 and C3 fields of view~\citep{Yashiro08_Flare-CME_corrs,salas15_Fx-Vcme_corr}. Besides, interactions between CMEs associated with $>$C-class flares have been shown to enhance SEP fluxes in $>$10\,MeV bands~\citep{gopal04_CME-CMEinter-SEP,gopal19_SGRE_CMEgrp}
Meanwhile, since the Fermi/LAT SGRE event lists cover only solar cycle 24 starting from 2011, this study focuses on CMEs associated with $>$C-class flares that occurred in major ARs between 2011 and 2019.}

{We define a major AR as an AR that produced more than one CME-associated major ($>$M-class) flare in its observed lifetime.
A catalog of AR sources of the CMEs associated with $>$C-class flares during 2011 - 2019 was made using data from the CDAW CME catalog version 2.0~\citep{cdawcme_catalogV2}\footnote{\href{https://cdaw.gsfc.nasa.gov/CME\_list/}{https://cdaw.gsfc.nasa.gov/CME\_list/}. The ARs that met the major AR criteria were then identified.}
For each major AR, we cataloged every associated CME, noting the SXR flare strength (\Fcme), \vcme, residual acceleration (\acme), onset time, and duration, forming the primary database for this study.
The CDAW CME catalog reports \vcme\ and \acme\ as the mean plane-of-sky speed and acceleration measured within the combined field of view of the LASCO C2 and C3 coronagraphs.
We express \Fcme\ on a linear scale where the GOES classes C1.0–C9.9 correspond to values 1.0–9.9, M1.0–M9.9 to 10.0–19.9, and $>$X1.0 to values above 20.0.}

The `major activity period' of each major AR is estimated as the interval between the first and last observed CMEs associated with $>$C-class flares. 
{The properties and occurrence of CMEs associated with $>$C-class flares during the major activity period were quantified for each AR using the parameters listed in Table~\ref{tab1:ARmetrics}, and their distributions were compared between `SGRE-producing' and `SGRE-lacking' AR populations.  
The distributions of \vcme\ and \acme\ of fast CMEs associated with $>$C-class flare in both major AR populations were analyzed and contrasted with those of SGRE-associated CMEs.

{After studying the CMEs and their general occurrence characteristics during major activity periods, we explored the temporal clustering of CME events during `fast CME epochs' in the major activity period of each major AR. We define a fast CME epoch as a 2-day interval centered on the fast CME onset time, and a `CME cluster' as a succession of CMEs associated with $>$C-class flares from an AR with waiting times less than a day.
CME clusters in the major ARs were identified, and their role in making co-temporal fast CMEs SGRE-associated was investigated. 
A detailed case study was performed on the CME clusters in two very active major ARs that produced multiple long-duration ($>$10\,h) SGRE (l-SGRE) events, to understand why fast CMEs in some clusters were SGRE-associated.

{ Table~\ref{tab:definitions}} summarizes all important definitions used in this work.
Previous studies that explored the extreme flare productivity of certain ARs had used convenient definitions to pick out super active ARs, or `super ARs'. Super ARs were first defined by \cite{Bai87_superARs} as ARs producing $>$4 major flares. 
Subsequent works refined the definition by adding criteria on photospheric magnetic field complexity, sunspot properties, radio flux, geomagnetic indices, and SEP flux levels, etc., but without focusing on CME-productivity~\citep{tian02_superARs,romano07_superARs,2011A&A...534A..47C,Wang13_QuasiHomoCMEs}.
Similarly, the common definition of a CME cluster does not constrain based on the AR source or associated flare strength of each CME, unlike the definition used in this work.
Our definitions of a major AR and a CME cluster are therefore distinct and tailored to strong CME-productive ARs and clustering of energetic events within them, which is relevant for SGRE production.}
\begin{table}[!htb]
\begin{tabular}{ll}
\hline
\textbf{Term} & \textbf{Definition} \\ 
\hline
 Fast CME& \begin{tabular}[c]{@{}l@{}}A CME with a sky-plane speed $>$1000\kms .\end{tabular}\\  
 Very fast CME& \begin{tabular}[c]{@{}l@{}}A CME with a sky-plane speed $>$1500\kms .\end{tabular}\\  
  SGRE-associated CME& \begin{tabular}[c]{@{}l@{}}A CME associated with an SGRE.\end{tabular}\\   
 \hline
 Major AR& \begin{tabular}[c]{@{}l@{}} An AR that produced more than one CME-associated \vspace{0.1cm}\\ major ($>$M-class) flare during its observed lifetime.\end{tabular}\vspace{0.1cm}\\
 Major activity period& \begin{tabular}[c]{@{}l@{}} The period between the first and the last observed \vspace{0.1cm} \\ CMEs associated with $>$C-class flares.\end{tabular}\vspace{0.1cm}\\
 SGRE-producing AR& \begin{tabular}[c]{@{}l@{}}An AR that produced $\geq$1 SGRE event in its observed lifetime.\end{tabular}\vspace{0.1cm}\\
 SGRE-lacking AR& \begin{tabular}[c]{@{}l@{}} An AR that produced no SGRE in its observed lifetime.\end{tabular}\\
 \hline
 fast CME epoch & \begin{tabular}[c]{@{}l@{}}A 2-day interval centered on the onset time of the fast CME.\end{tabular}\\
 CME cluster& \begin{tabular}[c]{@{}l@{}}A sequence of CMEs associated with $>$C-class flares \vspace{0.1cm}\\from an AR with waiting times less than a day. \end{tabular}\\
\hline
$>$3\,h SGRE & An SGRE event with a duration $>$3\,h\\
l-SGRE & Long-duration SGRE lasting for $>$10\,h\\
\hline
\end{tabular}%
\caption{{Definitions of various terms used in this work related to CMEs associated with $>$C-class flares, ARs, CME occurrence periods and SGRE events.}}
\label{tab:definitions}
\vspace{-8mm}
\end{table}

\section{Results and discussion} \label{results}
{We identified 76 major ARs between 2011 and 2019. All $>$3\,h SGRE events in the cycle were associated with 12 major ARs, 5 of which produced the 7 l-SGRE events in the cycle.
A table summarizing the properties of the major activity period and associated CMEs for every major AR is provided online\footnote{\href{https://cdaw.gsfc.nasa.gov/pub/atul/majorARs/MajorAR_CME-flareprops.html}{https://cdaw.gsfc.nasa.gov/pub/atul/majorARs/MajorAR\_CME-flareprops.html}}.
The CME clusters during the major activity periods were also identified in all major ARs, details of which are made available\footnote{\href{https://cdaw.gsfc.nasa.gov/pub/atul/majorARs}{https://cdaw.gsfc.nasa.gov/pub/atul/majorARs}}.
}

{This section is divided into four subsections following a top-down approach. 
Section~\ref{sec:AR_metrics_stats} discusses the results at the level of major activity periods and CMEs in general, in the major AR populations. Section~\ref{sec:fastCMEepoch} delves into the properties of fast CME epochs. Section~\ref{sec:CMEclusters} discusses the association of fast CME epochs and SGRE events with CME clusters. Finally, Sect.~\ref{sec:casestudy} presents a detailed case study exploring the role of CME-CME interactions during CME clusters in SGRE generation.}
\subsection{Properties of CMEs and major activity periods}~\label{sec:AR_metrics_stats}
\vspace{-1mm}
\begin{table}[!htb]
\begin{tabular}{cll}
\hline
\multirow{7}{*}{\textbf{\begin{tabular}[c]{@{}c@{}}CME\end{tabular}}} & \textbf{Parameter} & \textbf{Definition} \\ 
\hline
 & \Fcme & \begin{tabular}[c]{@{}l@{}}Strength of the associated flare on a scale \vspace{0.1cm}\\ linear with the GOES class (see, Sec 2)\end{tabular} \vspace{0.1cm}\\ 
 & \vcme & Sky-plane speed in the LASCO C2 field of view \vspace{0.1cm} \\  
 & \acme & \begin{tabular}[c]{@{}l@{}}Residual sky-plane acceleration in the \vspace{0.1cm}\\ LASCO C2 field of view\end{tabular} \vspace{0.1cm}\\ \hline
\multirow{5}{*}{\textbf{\begin{tabular}[c]{@{}c@{}}Major activity \vspace{0.1cm}\\ period\end{tabular}}} & \ardt & Duration of the major activity period\\  
 & \Ncme & No. of CMEs. \\  
 & \rcme & \Ncme/\ardt. \\  
 & \Fm & \Fcme\ of the strongest flare. \\  
 & \vcmem & \vcme\ of the fastest CME. \\ \hline
\end{tabular}%
\caption{Various parameters to characterize the strength of each CME event and major activity period in a major AR.}
\label{tab1:ARmetrics}
\vspace{-8mm}
\end{table}
{Table~\ref{tab1:ARmetrics} presents the various parameters of CMEs and major activity periods. A CME is characterized by \Fcme, \vcme, and \acme\ (see, Sec.~\ref{data}). Each major activity period is characterized by its duration (\ardt), number of CME events (\Ncme) and their rate (\rcme = \Ncme /\ardt), the maximum observed \vcme\ (\vcmem), and the maximum observed \Fcme\ (\Fm)}.

\begin{figure}[]    
\centerline{\includegraphics[width=\textwidth,height=\textheight]{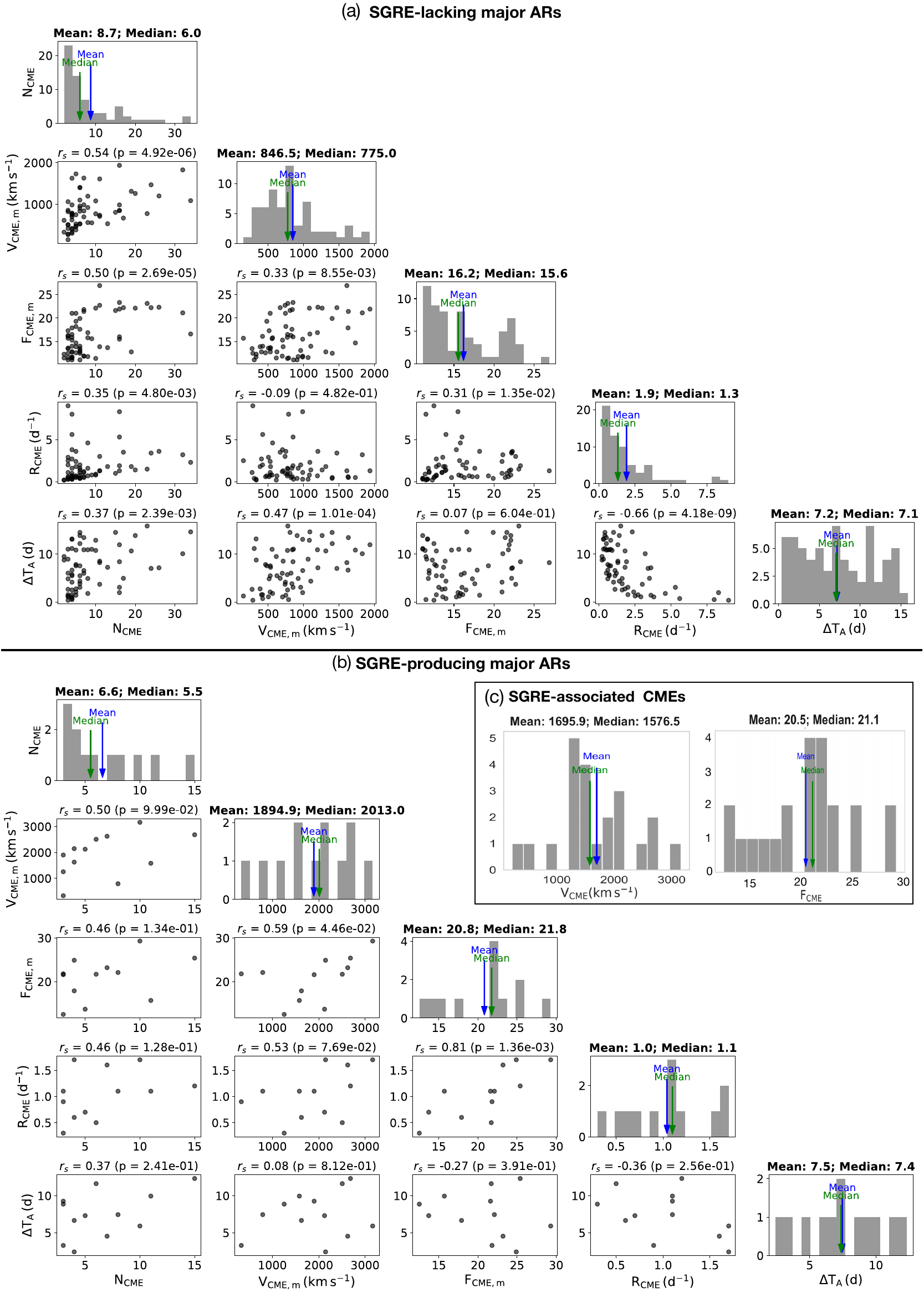}}
\caption{Major activity period characteristics. (a-b): Spearman cross-correlation matrix (\rs: correlation coefficient) with histograms for the various parameters of the major activity period (see Tab.~\ref{tab1:ARmetrics}) (c) Histogram of \vcme\ and \Fcme\ of SGRE-associated CMEs.}
\label{fig:majorARcomp}
\end{figure}

Figure~\ref{fig:majorARcomp} compares the distributions of and Spearman cross-correlations between the various parameters of the major activity period in the {SGRE-lacking and SGRE-producing} major AR populations. 
The median values of \Ncme, \rcme, \Fm, and \ardt\ are similar for the two populations, but the median \vcmem\ and \Fm\ are clearly higher (2013\kms; X1.8) for SGRE-producing ARs, compared to the other (775\kms; M5.8). 
Spearman correlation coefficient (\rs) between each parameter and the respective p value is mentioned in Fig.~\ref{fig:majorARcomp}. 
The apparent correlation between the \rcme\ and \ardt\ results from the definition of \rcme.
Meanwhile, based on p-value and correlation coefficient (\rs), significant (p$<$0.001) high correlations (\rs $\gtrsim$0.4) are found between parameter pairs (\ardt, \vcmem), (\ardt, \Ncme), (\vcmem, \Ncme), and (\Fm, \Ncme) for {SGRE-lacking major ARs}.
These parameter pairs remain significantly correlated even after including the SGRE-associated ARs, {although the \rs\ values} reduce by 12\%, 3\%, 20\%, and 12\%, respectively for each pair.
{These correlations may simply reflect that longer-lasting ($\approx$\ardt) and more CME-prolific ($\approx$\Ncme) major activity periods tend to host stronger CMEs that are accompanied by stronger flares. 
However, despite similar median values of \ardt, and \rcme, what distinguishes SGRE-producing ARs from the other population is their ability to generate exceptionally fast CMEs associated with stronger flares, as evidenced by their distributions for \vcmem\ and \Fm.
}

{Since CME kinematics clearly play a major role in SGRE generation, we explored the distribution of \vcme\ and  \acme\ for all fast CMEs in major AR populations, compared with the SGRE-associated CMEs (see Fig.~\ref{fig:cmecomp}).
}
\begin{figure}[]  
\centerline{\includegraphics[width=\textwidth,height=0.47\textheight]{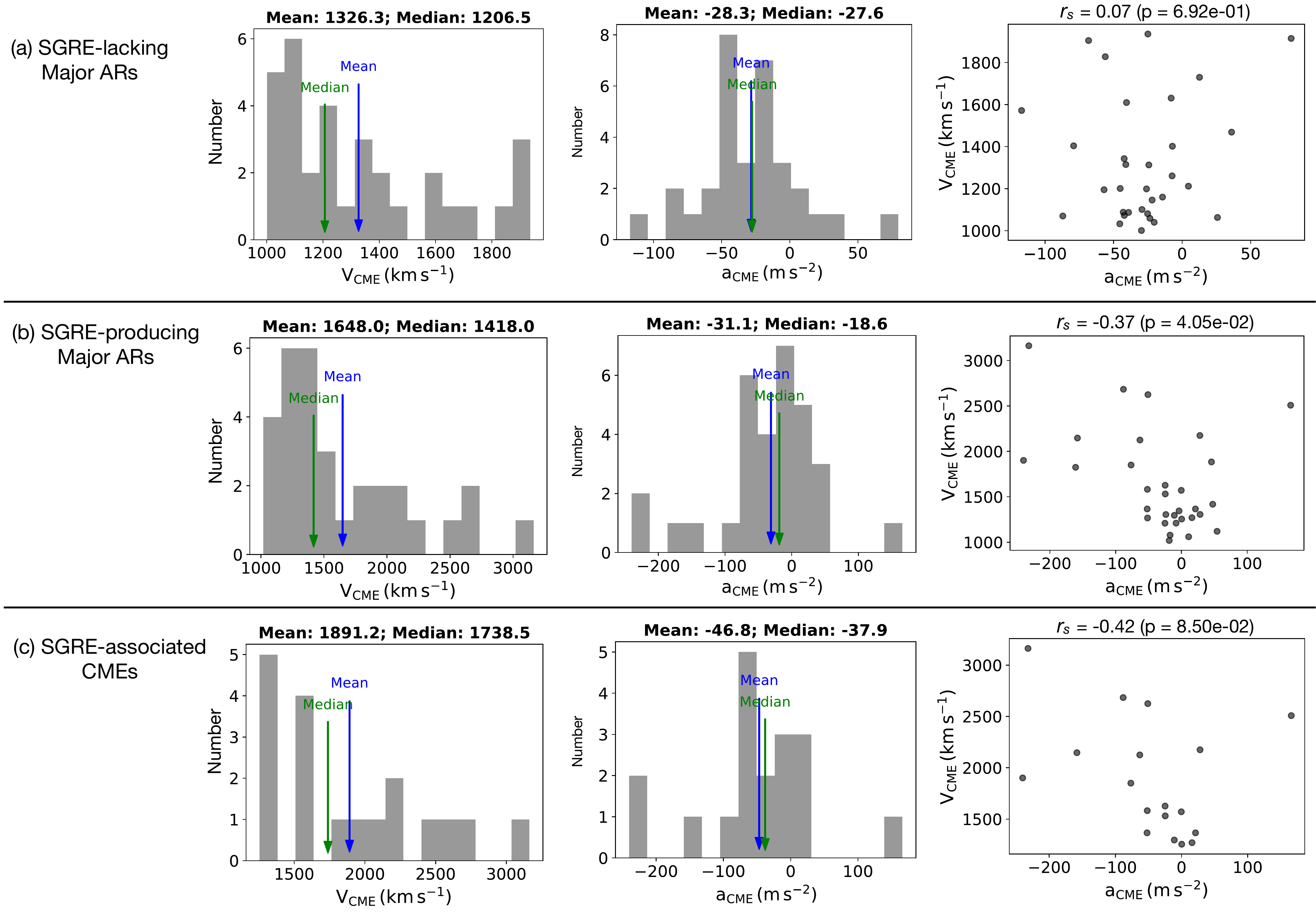}}
\caption{Properties of fast CMEs in SGRE-lacking (a) and SGRE-producing (b) major ARs and those associated with SGRE events (c). Distributions of \vcme\ and \acme, and the Spearman cross-correlation between \vcme\ and \acme\ are shown.}
\label{fig:cmecomp}
\end{figure}
{Of the major AR populations, SGRE-producing ARs generate relatively faster CMEs. The SGRE-associated CMEs form a much faster population of CMEs with very high median residual deceleration (-\acme) within their host ARs.
}
The \acme\ and \vcme\ show a significant anti-correlation for fast CMEs in SGRE-producing ARs and for SGRE-associated CMEs. 
This could be because faster CMEs face a high residual deceleration due to a higher solar wind drag force ($\propto$\vcme$^{2}$)~\citep[see,][for an overview]{chen11_CMEmodelsRev}. 
This effect was previously reported for samples of mostly isolated CME events whose evolving structures are well discernible in coronagraphs~\citep{Gopal00_aCME-VcmeCorr_arrivtimPred,Gopal01_aCME-VcmeCorr_arrivtimPred,2001JGR...10629219G}.
However, the lack of a \vcme\ -\acme\ anti-correlation in the fast CMEs of SGRE-lacking ARs is intriguing, which might be related to the mean ambient conditions during fast CME epochs. {The correlation between \vcme\ and \acme\ is clearly pronounced for very fast ($>$1500\kms) CMEs, whose dynamics will be relatively less affected by the structures in the propagating medium or near simultaneous eruptions.
The following subsection will characterize the fast CME epochs in major ARs, and compare with the SGRE-associated CME epochs.}

\subsection{Activity of CMEs during fast CME epochs}~\label{sec:fastCMEepoch}
{Of the 64 }SGRE-lacking major ARs, 17 produced at least one fast CME, and 6 ARs produced at least one very fast CME. Meanwhile, out of the 12 SGRE-producing ARs, 9 hosted very fast CMEs, while one produced a 1200\kms\ CME and two outlier ARs produced $<$1000\kms\ CMEs associated with SGRE events lasting for $\lesssim$5\,h.
\begin{table}[!htb]
\begin{tabular}{ll}
\hline
\textbf{Parameter} & \textbf{Definition} \\ 
\hline
 \twb (\twf) & \begin{tabular}[c]{@{}l@{}}Waiting time of the previous (following) CME\end{tabular}\\ 
 \rcmeb (\rcmef) & \begin{tabular}[c]{@{}l@{}}CME rate in the period before (after) the event onset\end{tabular}\\  
 \vcmerb (\vcmerf) & \begin{tabular}[c]{@{}l@{}}Relative sky speed of the preceding (following) CME\end{tabular} \\
 \lcb (\lcf) & Interaction distance parameter with the previous (following) CME\\  \hline
\end{tabular}%
\caption{Various parameters to characterize the CME activity in the periods just before (following) each fast CME event. The parameters are defined in the same sense for both periods, with subscripts `b' and `f' denoting the periods before and following the fast CME of interest.}
\label{tab2:Epochmetrics}
\vspace{-8mm}
\end{table}

\begin{figure}[!htb]   
\centerline{\includegraphics[width=\textwidth,height=0.4\textheight]{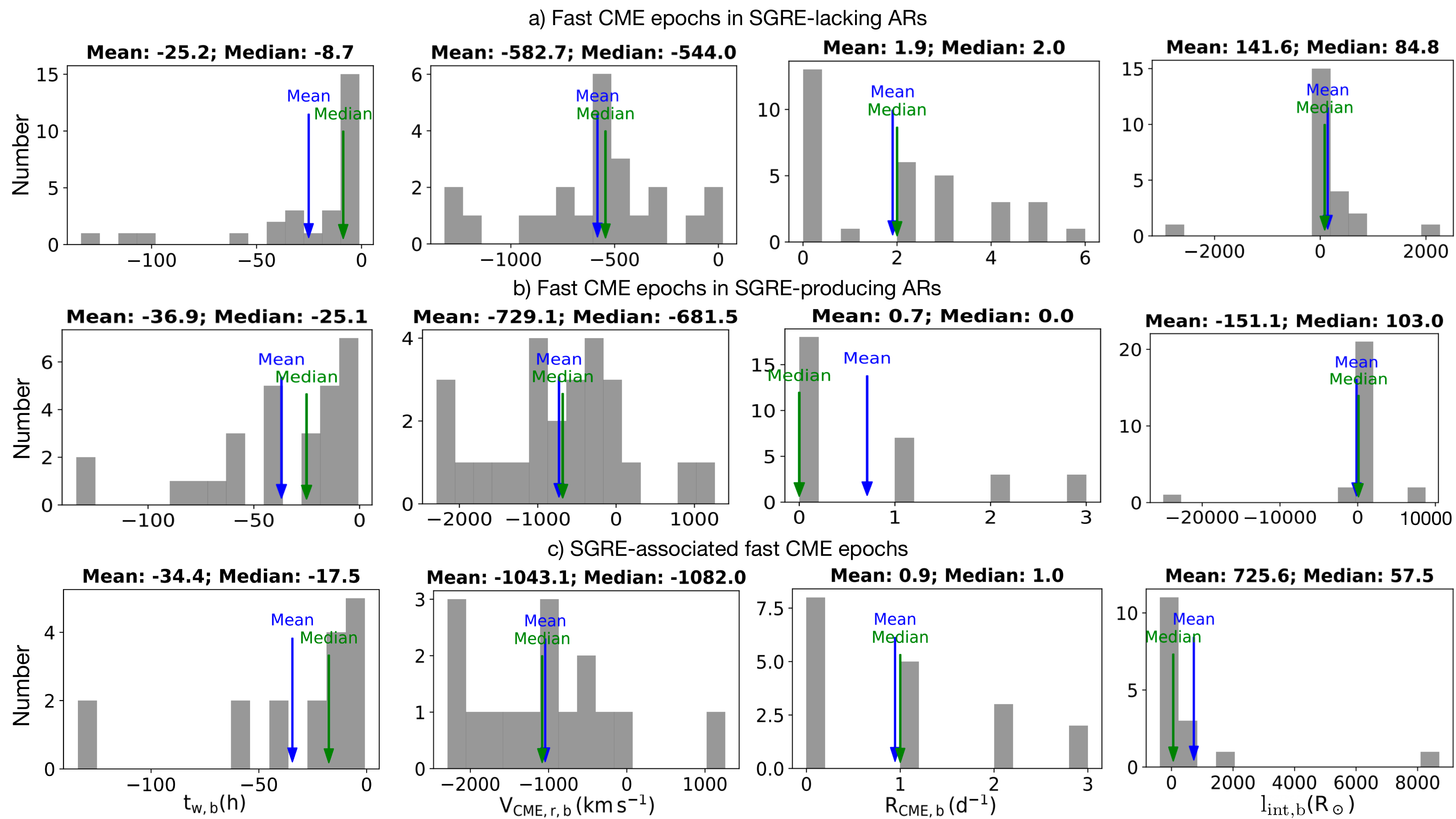}}
\caption{Statistics of the parameters of CME activity in the period just before the fast CME onset within fast CME epochs; (a) in SGRE-lacking major ARs, (b) in SGRE-producing major ARs and (c) associated with SGRE events.}
\label{fig:fastcmeb}
\end{figure}
\begin{figure}[]     
\centerline{\includegraphics[width=\textwidth,height=0.4\textheight]{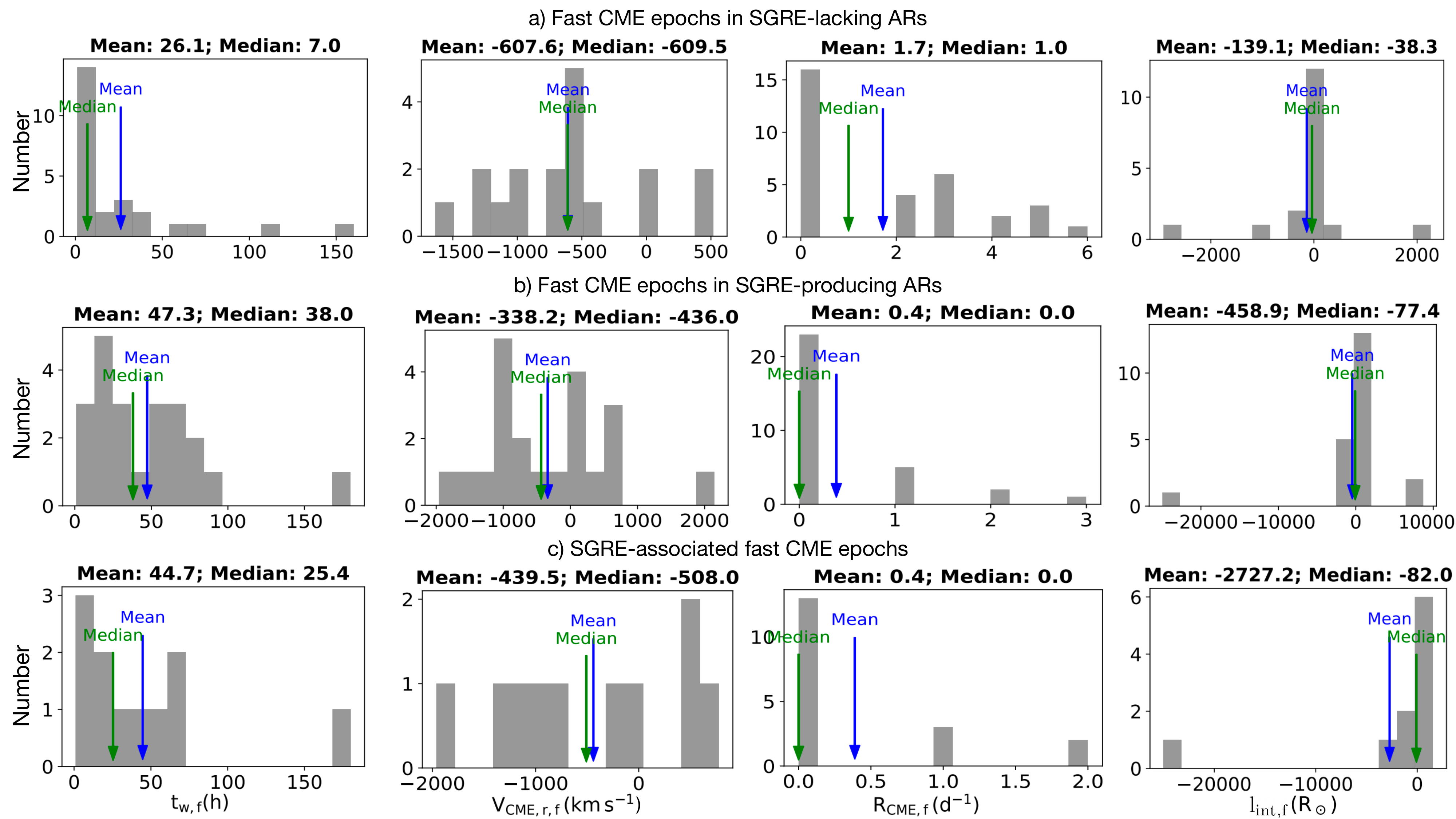}}
\caption{Same as Fig.~\ref{fig:fastcmeb}, but characterizing the period following the fast CME onset within fast CME epochs.}
\label{fig:fastcmef}
\end{figure}
To characterize each fast CME epoch, {we define two sets of parameters for the 1-hour period just before and following the event onset, which are summarized in Tab.~\ref{tab2:Epochmetrics}. The various parameters include the waiting time for the immediately preceding (\twb) and following (\twf) CMEs, the CME rates (in d$^{-1}$) before (\rcmeb) and after (\rcmef) the fast CME onset, the relative speed of the immediately preceding (\vcmerb) and following (\vcmerf) CMEs with respect to the fast CME of interest, and the typical interaction distance of the fast CME with the immediately preceding (\lcb) and following (\lcf) CME.}
A negative value of \vcmerb (\vcmerf) implies a relatively slower speed for the preceding (following) CME.
{The \lcb\ (\lcf) value is computed based on the relative speed and onset time difference of the preceding (following) CME with respect to the fast CME of interest.
Since the CMEs considered to compute \lcb\ (\lcf) occur in the same AR, their sky-plane speeds would have very similar projection effects.
However, since the details of the CME dynamics are not considered, the estimates of the interaction distance are nominal with lower positive values of \lcb\ (\lcf) indicating higher chances of CME-CME interaction.}
A negative \lcf\ (\lcb) implies that the CME of interest would not interact with the following (preceding) CME. {We remind the reader that only CMEs associated with $>$C-class flares are considered in this study.}

{Figures ~\ref{fig:fastcmeb} and ~\ref{fig:fastcmef} show the histograms of all parameters in Tab.~\ref{tab2:Epochmetrics} characterizing the CME activity during the fast CME epochs of SGRE-producing and SGRE-lacking major ARs. For comparison, statistics for SGRE-associated fast CME epochs are shown separately.
Since parameter distributions are generally asymmetric and have outliers, we take the median value of each parameter as its representative estimate.}
Note that \twb\ is negative while \twf\ is positive because waiting times are measured from the epoch of the fast CME. We will hence use their absolute values for a comparative analysis of the periods preceding and following the fast CME onset time.
The absolute values of \twb\ and \twf\ are low, while \rcmef\ and \rcmeb\ are high for SGRE-lacking ARs, compared to SGRE-producing ARs. 
This could indicate higher chances for CME-CME interactions in SGRE-lacking ARs.
However, the median absolute \vcmerb\ is higher for SGRE-producing ARs.
Since fast CMEs in SGRE-producing ARs are on average faster than those in SGRE-lacking ARs (see, Fig.~\ref{fig:majorARcomp}), the higher \vcmerb\ implies that the interactions of their fast CMEs with the preceding CMEs could be much stronger.
The interaction distance is another important parameter because CMEs expand while propagating away from the Sun, causing plasma and magnetic field energy densities to decrease. Hence, CME-CME interactions at lower distances are more energetic with higher particle acceleration potential compared to those occurring at larger distances. 
The median \lcb\ of SGRE-producing ARs is higher compared to SGRE-lacking ARs. 
Meanwhile, negative median values of \lcf\ for the three populations of fast CME epochs {are likely due to the low probability of the occurrence of multiple fast CMEs in short waiting times. This will result in a negative median value for \vcmerf\ and in turn \lcf.}

Next, we investigate {CME activity in SGRE-associated CME epochs and compare with the general results for fast CME epochs in the two major AR populations.} 
SGRE-associated CME epochs show significantly lower \lcb\ and higher \vcmerb\ than {typical fast CME epochs in major ARs. This suggests higher chances of stronger CME-CME interactions during SGRE-associated CME epochs.} 
In addition, the {relatively high median value of \rcmeb\ for SGRE-associated CME epochs compared to a typical fast CME epoch in SGRE-producing ARs suggest that SGRE-associated CME epochs are distinctive periods of high CME activity in their host ARs.}
To summarize the general CME {activity characteristics during fast CME epochs,
\begin{enumerate}
    \item Fast CME epochs in SGRE-lacking ARs show higher chances of CME-CME interactions. 
    \item The higher median values for \vcme\ and \vcmerb\ for fast CMEs and their epochs respectively in SGRE-producing ARs indicate the chances of relatively stronger CME-CME interactions in these ARs.
    \item SGRE-associated CME epochs are periods of higher CME activity in their host ARs, with very high chances of strong CME-CME interactions at low interplanetary space heights.
\end{enumerate}
}
Though the epochs of SGRE-associated CMEs and fast CMEs in SGRE-lacking ARs show high \rcmeb, the high \vcmerb\ and low \lcb\ clearly distinguish the former from the latter. 
{A robust investigation of CME-CME interactions during fast CME epochs requires systematic identification and characterization of CME clusters and co-temporal fast CMEs in major ARs.}
\subsection{CME clusters in major ARs}~\label{sec:CMEclusters}
We identified 94 CME clusters across multiple major ARs. The plots of CME occurrences during the major activity period of each major AR with clusters marked are available online, along with the properties of the CMEs and associated flares\footnote{\href{https://cdaw.gsfc.nasa.gov/pub/atul/majorARs/CME_clusters}{https://cdaw.gsfc.nasa.gov/pub/atul/majorARs/CME\_clusters}}. 
Of the identified clusters, 19 found in 16 major ARs had at least one fast CME, and 12 found in 8 ARs had at least one very fast CME. 
We find that 12 out of 22 (54.5\%) SGRE events and 5 out of 7 (71.4\%) l-SGRE events occurred in CME clusters. Exploring all these clusters individually in detail is beyond the scope of this paper.
{Therefore, we resort to a case study of CME clusters in very active major ARs that produced multiple l-SGRE events.
ARs 11429 and 12673 are the only such ARs, with some l-SGRE events associated with CME clusters.} 

\subsection{{Case study: CME clusters in very active major ARs}}~\label{sec:casestudy}
\begin{table}[!htb]
\begin{tabular}{clcccccl}
\hline
AR & Event date & Time & Dur & GOES & \vcme & \vsp & CME\\
     &      & UT & h & class & km\,s$^{-1}$ & km\,s$^{-1}$ & cluster \\

\hline
 & 2012-03-05 & 04:09 & 4.25 & X1.1 & 1531 &1627 & Yes \\ 
 & \textbf{2012-03-07} & \textbf{00:24} & \textbf{20.47} & \textbf{X5.4} & \textbf{2684}& \textbf{3146} & \textbf{Yes} \\ 
 & 2012-03-09 & 03:53 & 8.85 & M6.3 & 950 & 1229 & No \\ 
\multirow{-4}{*}{11429} & 2012-03-10 & 17:44 & 11.62 & M8.4 & 1296 & 1638 & No \\ \hline
\multicolumn{1}{l}{} & \textbf{2017-09-06} & \textbf{12:02} & \textbf{18.43} & \textbf{X9.3} & \textbf{1571} & \textbf{1819} & \textbf{Yes} \\ 
\multicolumn{1}{l}{\multirow{-2}{*}{12673}} & {2017-09-10} & {16:06} & {15.18} & {X8.2} & {3163} & {3163} & \textbf{Yes} \\ \hline
\end{tabular}%
\caption{SGRE events produced by ARs 11429 and 12673.}
\label{tab:SGREARs}
\end{table}
\begin{figure}    
\centerline{\includegraphics[width=\textwidth,height=0.24\textheight]{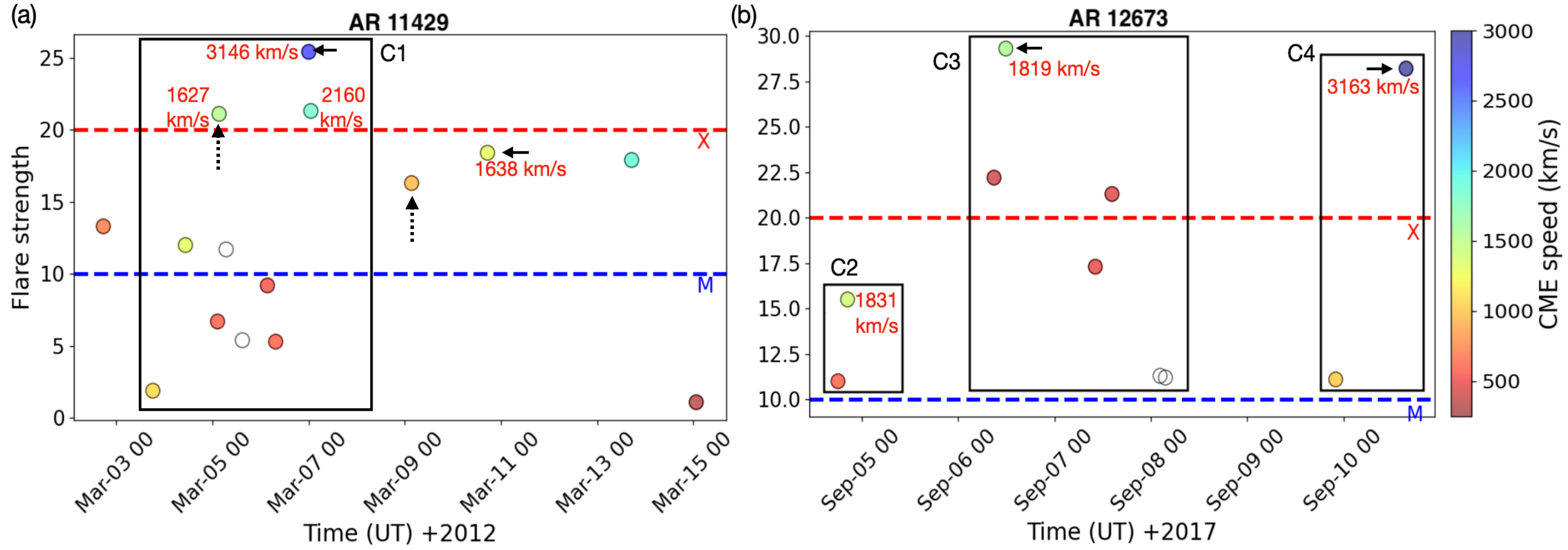}}
\caption{CMEs produced during the major active periods in the selected major ARs. Boxes identify CME clusters (C1 - C4) and horizontal solid arrows mark the l-SGRE-associated CMEs. Vertical dotted arrows mark the other SGRE-associated CMEs. The \vsp\ of all very fast CMEs are mentioned in red text. Label colors provide \vcme.}
\label{fig:selectARs}
\end{figure}

Table~\ref{tab:SGREARs} summarizes the properties of the recorded SGRE events in {ARs 11429 and 12673} and the associated CMEs. Since \vcme\ {is the projected speed in the sky-plane}, the true space speed (\vsp) of every SGRE-associated CME was estimated by fitting a cone CME model~\citep{xie06_conemodel}.
The columns specify the event date, time, duration, associated flare class, CME speeds, and if the SGRE-associated CME occurred in a cluster. The rows in bold highlight the l-SGRE events that occurred in a cluster. All SGRE-associated CMEs were halo CMEs.
Figure~\ref{fig:selectARs} shows the various CME clusters during the major activity period of the selected ARs. The solid arrows point to l-SGRE-associated CMEs, {when the dotted arrows point to other SGRE-associated CMEs}. \vsp\ values of the fast CMEs are shown in red text.
Note that cluster C2 in AR 12673 did not produce even a 3-hour-long SGRE, despite having a very fast CME with a \vsp\ similar to that {in }cluster C3. The following sub-subsections will explore these ARs and clusters in detail.
\subsubsection{AR 11429}\label{sec:AR11429}
Cluster C1 produced by the AR lasted from 2012-03-03 18:36 to 2012-03-07 01:30 UT. {Two SGRE events that lasted for 20.47\,h (l-SGRE) and 4.25\,h occurred during the cluster. } 
{AR moved from 83$^\circ$ - 26$^\circ$ east longitude (latitude: 16$^\circ$N) during the cluster period.} 
Figure~\ref{fig:AR11429_cluster} shows the SEP flux from GOES in the top panel and the CME height-time plots for all CMEs that occurred {during cluster C1}, and could be robustly tracked. The data comes from the height-time measurements in the CDAW CME catalog. The height-time {data points are colored based on the CME propagation direction, and the slope of the data gives \vcme.} Green dotted arrows highlight the SGRE-associated CMEs. The box identifies CME-CME interactions, with black arrows pointing at the interacting CMEs. 
\begin{figure}    
\centerline{\includegraphics[width=\textwidth,height=0.25\textheight]{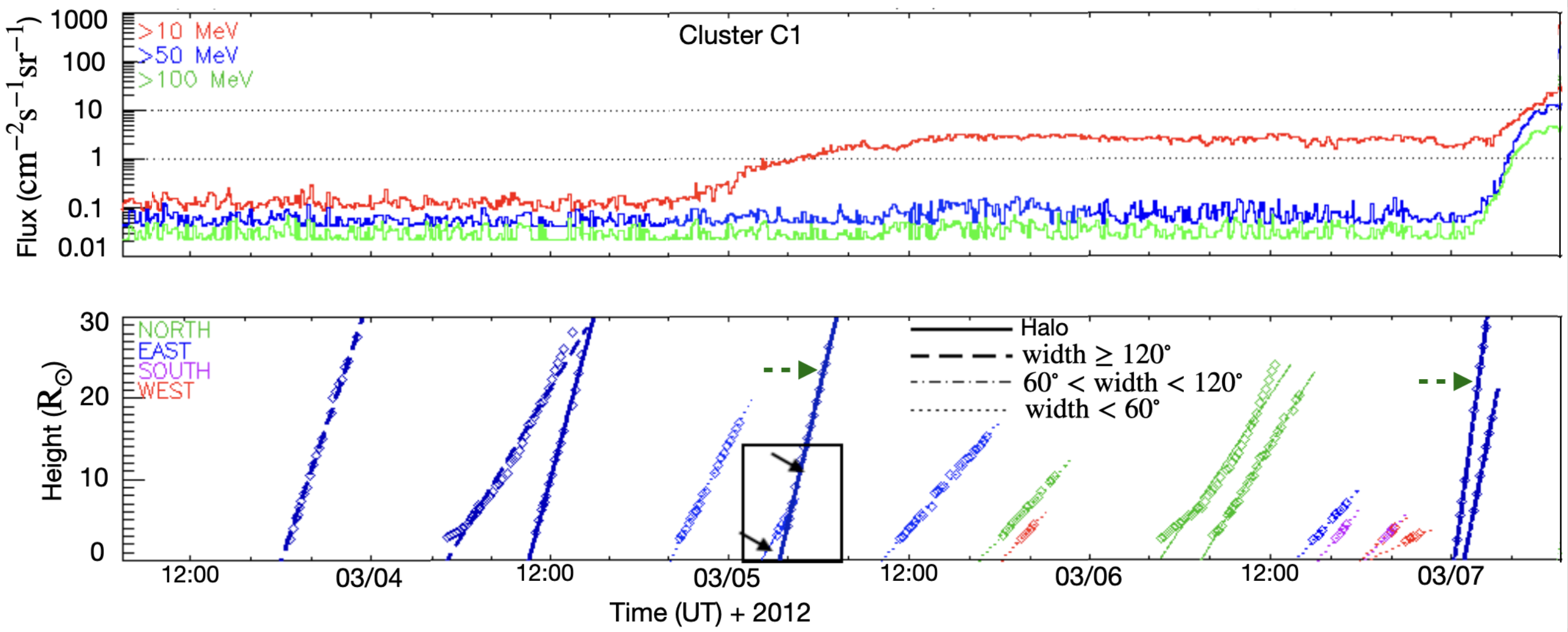}}
\caption{Activity around the period of cluster C1. Top panel shows SEP flux from GOES, while the bottom panel shows the CME height-time measurements marked by diamonds with best-fit lines drawn in different line styles depending on the CME width (see, plot legend). The line and marker colors denote the CME propagation direction. Eastern and northern CMEs are associated with AR 11429. Interacting CMEs from AR 11429 are boxed and marked by black arrows. SGRE-associated CMEs are marked by green dotted arrows.}
\label{fig:AR11429_cluster}
\end{figure}
The interaction involving a fast CME of \vsp\ = 1627\kms\ and \acme=-24.6\msq\ was associated {with an SGRE that lasted for 4.25\,h.
The interaction occurred at a short heliocentric distance $<$10\,\Rsun\ {with a waiting time and a speed of -0.78\,h and -937\kms, respectively, for the trailing CME relative to the fast CME. These values indicate that the trailing CME was relatively very weak and occurred temporally quite close to the fast CME.}
Meanwhile, the l-SGRE event on 2012-03-07, associated with a CME of \vsp= 3146\kms, show no sign of interactions.} 
The {two east-directed close-by CMEs on 2012-03-04 are not considered because the earlier of the two events is associated with an AR at N32E87, which is much further away from AR 11429 (N19E61) that produced the later CME.
}

Figure~\ref{fig:AR11429_fig2} explores the epochs of the other two SGRE-associated CMEs from AR 11429, {that were not associated with any cluster.} The 2012-03-09 03:53 UT event had poor-quality data in LASCO and STEREO A, making the analysis difficult. 
However, {the CME associated with the l-SGRE event on 2012-03-10 17:44 UT underwent an interaction with a CME from the nearby AR 11430 at a height within 10\,\Rsun.} The {waiting time and speed of the trailing CME relative to the fast CME in the interaction are -1.5\,h and -777\kms\ respectively.} {AR 11430 was located $\sim$12$^\circ$ west of AR 11429.} 
\begin{figure}    
\centerline{\includegraphics[width=0.65\textwidth,height=0.25\textheight]{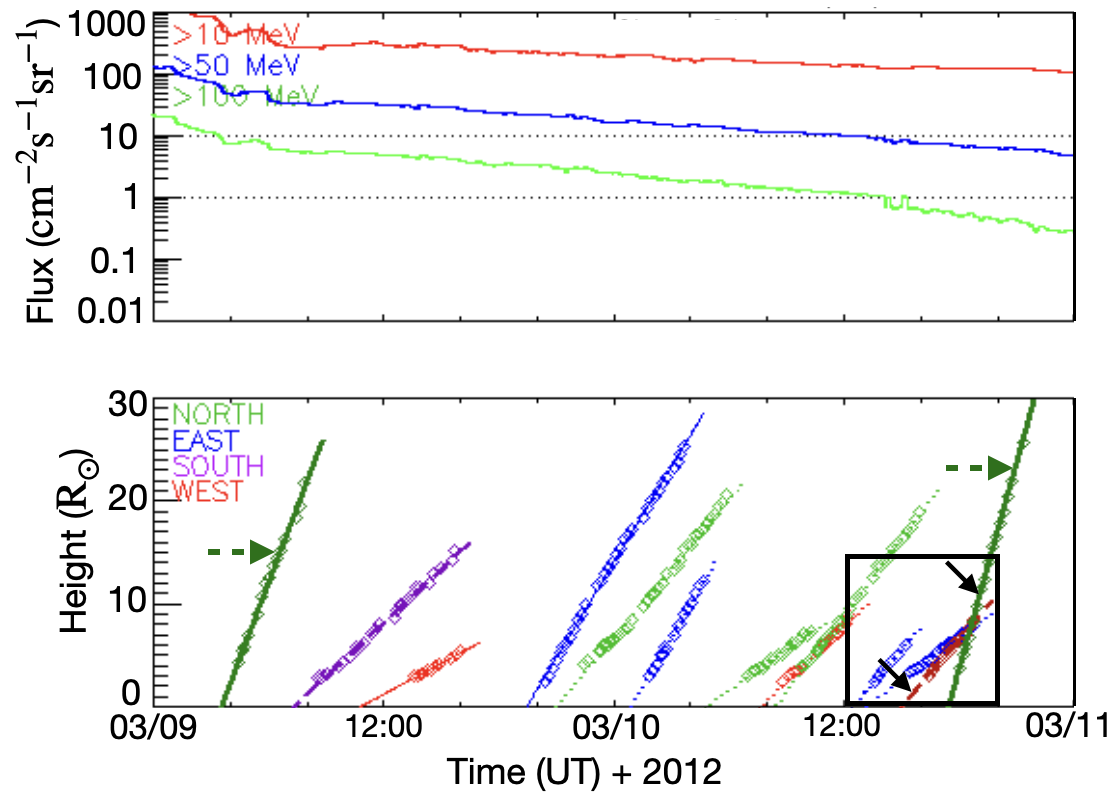}}
\caption{CME activity during {SGRE events in AR 11429 (N17W24) unrelated to CME clusters.} Figure panels and labels have the same meaning as in Fig.~\ref{fig:AR11429_cluster}. {The CME associated with the SGRE on 2012-03-10 17:44 UT interacted with an earlier CME from AR 11430 (N18W36).}}
\label{fig:AR11429_fig2}
\end{figure}

\subsubsection{AR 12673}
The AR produced 3 clusters, C2 - C4. C2 had two CMEs that were associated M-class flares on 2017-09-04 at 19:00 and 20:36 UT (See, Fig.~\ref{fig:selectARs}b). The latter event was a very fast (1831\kms) halo CME. The AR location was S08W11. After a short break, C3 and C4 occurred lasting from 2017-09-06 09:48 to 2017-09-08 04:17 UT and 2017-09-09 23:12 to 2017-09-10 16:00 UT respectively. {The location of the AR} varied from S08W11 to S08W92 during the period.
\begin{figure}    
\centerline{\includegraphics[width=0.65\textwidth,height=0.45\textheight]{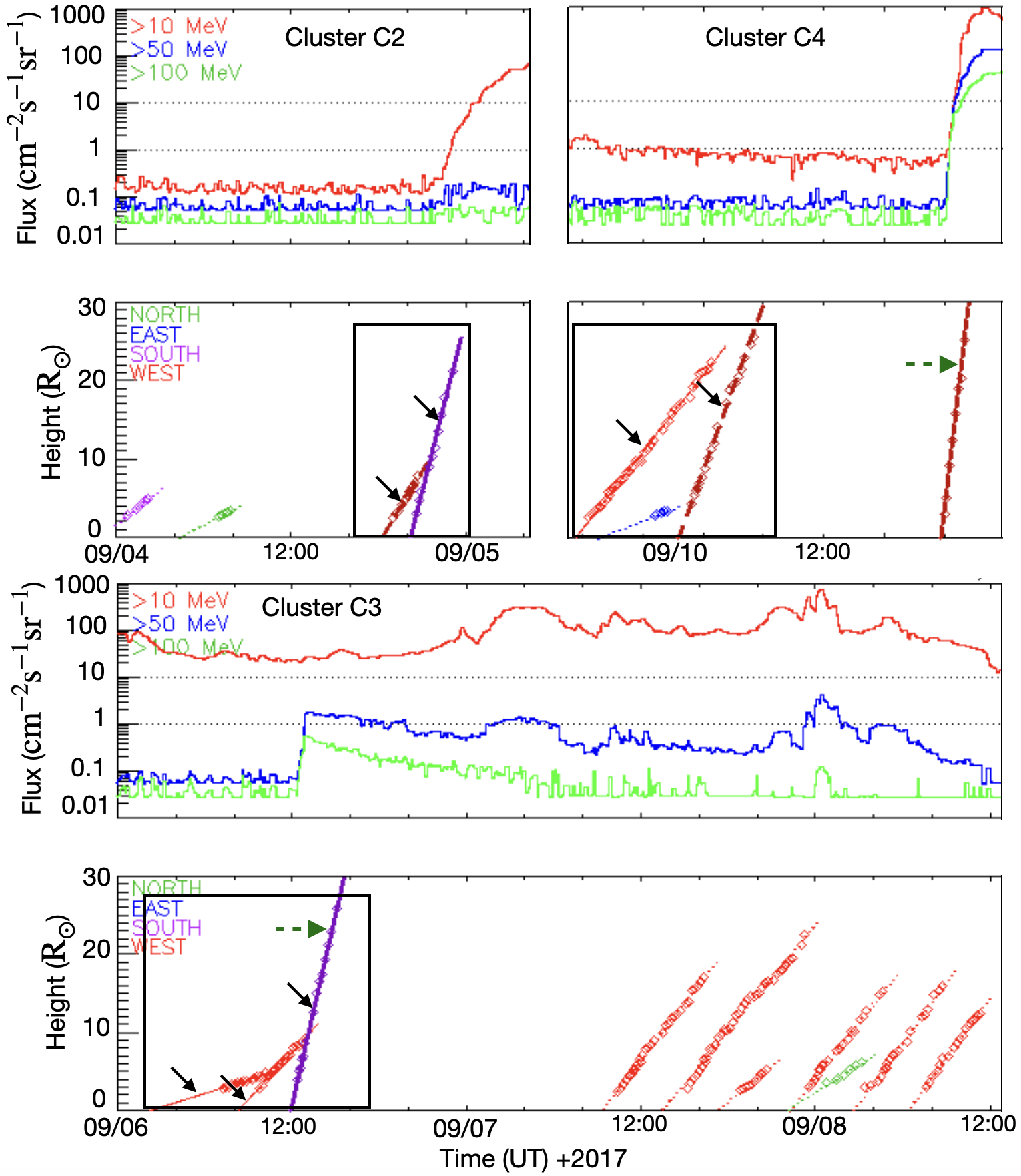}}
\caption{CME clusters in AR 12673. Markers, line styles and labels have the same meaning as in Fig.~\ref{fig:AR11429_cluster}. }
\label{fig:AR12673_clusters}
\end{figure}
\begin{figure}    
\centerline{\includegraphics[width=\textwidth,height=0.32\textheight]{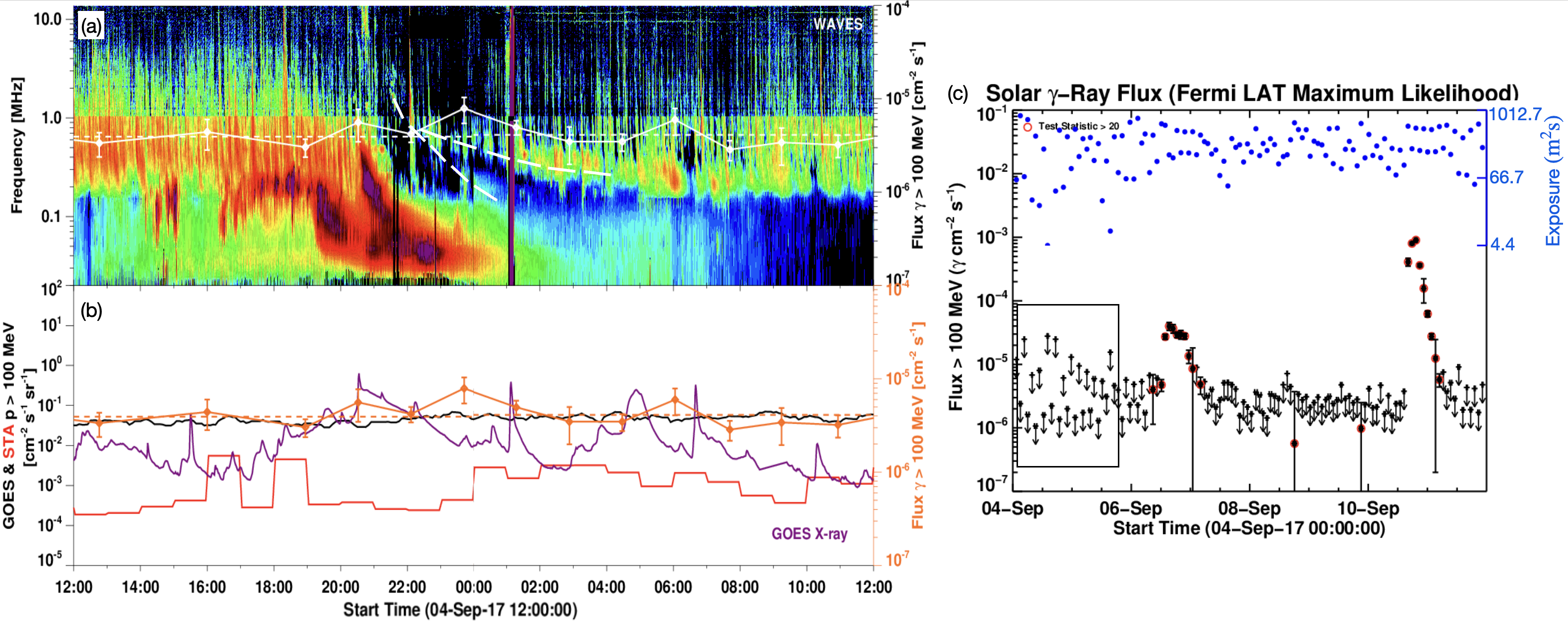}}
\caption{Activity during cluster C2. (a): DH dynamic spectrum showing the type II bursts marked by lines and the spike in $\gamma$-ray flux. (b) GOES X-ray flux (purple), and $>$100\,MeV proton fluxes from GOES (black) and STEREO A (red) are shown along with $\gamma$-ray flux in orange. (s) The LAT exposures and $\gamma$-ray fluxes from the maximum likelihood method are shown. Red points mark the two l-SGRE events that happened in AR 12673. The box region highlights the period of C2.}
\label{fig:C2_prop}
\end{figure}
Figure~\ref{fig:AR12673_clusters} shows an analysis of the clusters similar to that in Fig.~\ref{fig:AR11429_cluster}. {We first consider the case of SGRE-associated clusters, C3 and C4.}
In the case of C3 where a $<$2000\kms\ CME produced an l-SGRE, we find clear evidence for multi-step CME interactions involving 3 CMEs, all within 10\,\Rsun. The relative CME speeds and waiting times of the two trailing CMEs relative to the fast CME are (-1160\kms, -2.5\,h) and (-1015\kms, -9.6\,h).
{Cluster C4 has a case of CME-CME interaction between fast CMEs from AR 12673 marked by the black box. The interaction that occurred at a height above 20\,\Rsun\ did not produce an SGRE. {The trailing CME had a speed of about -500\kms, with a high waiting time of -5.5\,h with respect to the fast CME in the interaction event.} Meanwhile, the l-SGRE-associated CME on 2017-09-10 16:06 UT had a \vsp\ of 3163\kms\ with no sign of CME-CME interactions.} 

Now we consider the case of C2, which did not produce an SGRE despite having a CME-CME interaction involving a fast CME of speed similar to that in C3. 
The waiting time and speed of the trailing CME were -1.5\,h and -821\kms\ respectively, relative to the fast CME in the interaction.
The DH dynamic spectrum clearly shows type II radio bursts associated with the CMEs, highlighted by dashed white lines in Fig.~\ref{fig:C2_prop}. The overlaid $\gamma$-ray light curve shows a minor increase above the mean background level, observed after 2017-09-04 22:00 UT when the CME interaction occurred (see, Fig.~\ref{fig:AR12673_clusters}).
Figure~\ref{fig:C2_prop}c shows the LAT Maximum likelihood light curve during the major activity period of AR 12673. The SGRE events during clusters C3 and C4 are highlighted in red circles. The period of C2 is boxed. The flux estimation uncertainties during the C2 are as high as the detected SGRE fluxes for the event during C3. 
FERMI/LAT exposures are also very low during the C2 period. {Poor exposure coupled with high flux uncertainties could have hampered the detection of an SGRE in the C2 period, even if it had had a flux level similar to the event in C3.}


\section{Conclusion} 
\label{conclusion} 
This study investigates the role of CME clustering and CME-CME interactions in $>$3\,h SGRE events. We focus on major ARs, defined as active regions that produced more than one CME associated with a major ($>$M-class) flare during its recorded lifetime. {A CME cluster is defined as a series of CMEs with waiting times of less than one day. The study focuses only on CMEs associated with $>$C-class flares produced by major ARs.}

We identified 76 major ARs in cycle 24 since 2011, {using the elaborate database of CMEs, flares, and associated ARs compiled by NASA's CDAW data center.} Of these major ARs, 12 produced all SGRE events in the FERMI/LAT event lists. The properties of all major ARs are available online. We characterized the major activity period, defined as the period when an AR produced CMEs associated with flares $>$C-class, of all 76 ARs {using }various parameters. 
The parameters include the duration (\ardt), the fastest CME speed (\vcmem), the strongest flare strength (\Fm), the CME count (\Ncme) and the average CME rate (\rcme = \Ncme/\ardt) associated with the major activity period.
{Our analysis reveals significant correlations between (\vcmem, \Ncme), (\Fm, \Ncme), and (\ardt, \vcmem). This might be a reflection of the fact that longer major activity periods and higher \Ncme\ increase the chances for occurrence of faster CMEs.
Since SGRE events are closely associated with fast CMEs ($>$ 1000\kms), we also examined the distribution of their {sky-plane} velocity (\vcme) and residual acceleration (\acme) within the combined field of view of the LASCO C2 and C3 coronagraphs.
We conclude the following about CME activity during the major activity periods of the two major AR populations.
\begin{itemize}
	\item SGRE-lacking and SGRE-producing ARs have similar \ardt, and \rcme. 
	\item The strongest events in SGRE-producing ARs recorded a much higher median \vcmem\ (2013\kms) and flare class (X1.8), compared to SGRE-lacking ARs (775\kms, M5.8).
        \item  Fast CMEs in SGRE-producing ARs have a higher median value of \vcme\ (1418\kms) than in SGRE-lacking ARs (1206.5\kms).
	\item SGRE-associated CMEs formed a distinct population with significantly higher median \vcme\ (1738.5\kms) and residual deceleration (\acme = -34.9\msq) within host ARs.        
\end{itemize}
The SGRE-productivity of a major AR is determined by its ability to produce a population of very fast CMEs during its major activity period, independent of the extent of the major activity period or the mean CME rate.}

{We investigated the clustering and interaction of CMEs during fast CME epochs to explore their potential role in SGRE generation. A fast CME epoch is defined as a 2-hour interval centered on the fast CME onset time. For each fast CME epoch, we estimated the waiting times and relative speed of the immediately preceding and following CMEs. We also computed the CME rates before and after the fast CME onset and interaction distances of the fast CME with the immediately preceding and following CMEs. All fast CME epochs in major ARs were characterized using these parameters.
SGRE-associated CME epochs showed very low median values for the interaction distance and CME waiting times, with a high median value for the relative speed with respect to the preceding CME, compared to typical fast CME epochs in major ARs. In addition, SGRE-associated CMEs occurred during periods of enhanced CME rates in their host ARs. Hence, an SGRE-associated CME epoch has the highest probability of a strong CME-CME interaction compared to other fast CME epochs. 
}
In addition, we identified 96 CME clusters in all major ARs. 
Twelve of 22 (54.5\%) SGRE events and five of 7 (71.4\%) long-duration SGRE (l-SGRE) events occurred in CME clusters. Our findings suggest that, like the SGRE-associated fast CMEs, {their occurrence epochs also form a distinctive group within their major ARs characterized by higher odds of association to CME clusters and CME-CME interaction.}

For a detailed insight, we performed a case study of CME clusters within two very active major ARs, 11429 and 1267, which together hosted four CME clusters and multiple l-SGRE events. All SGRE-associated CMEs in these ARs were halo CMEs with true space speeds ranging from 1627 to 3163\kms.
AR 1267 was particularly notable, as it had three CME clusters, two of which produced l-SGRE events.
{We found that all fast CMEs with a speed below 2000\kms\ that are associated with SGRE events had interactions with previous CMEs within distances below 10\,\Rsun. However, l-SGRE-associated CMEs with speeds $>$3000\kms\ did not undergo CME-CME interactions.}

{These findings emphasize the crucial role of CME clusters and associated CME-CME interactions in the SGRE productivity of fast CMEs with speeds below 2000\kms. However, exceptionally fast CMEs (\vcme $>$ 3000\kms) are capable of triggering l-SGRE events independently. We find that the ability of a major AR to produce very fast CMEs in clusters strongly determines its SGRE productivity.}

\begin{acks}
This work was initiated by the international workshop entitled “Origin of High-Energy Protons Responsible for Late-Phase Pion-Decay Gamma-Ray Continuum from the Sun” supported by the Institute for Space-Earth Environmental Research, Nagoya University.
AM and NG are partly supported by NASA’s STEREO project, LWS program and Internal Scientist Funding Model (ISFM). PM, SA, and SY were partially supported by NSF grant, AGS-2043131. We thank the CDAW team for maintaining an up-to-date catalog of solar CMEs detected by the Large Angle and Spectrometric Coronagraph
(LASCO) on board the Solar and Heliospheric Observatory (SOHO) mission. The authors also thank the Fermi/LAT team for the gamma-ray analysis codes and catalogs. 
\end{acks}









\begin{ethics}
\begin{conflict}
The authors declare that they have no conflicts of interest that could potentially bias the work or influence the publication process.
\end{conflict}
\end{ethics}

\bibliographystyle{spr-mp-sola}
\bibliography{paper}  


\end{document}